# Pressure enhanced interplay among lattice, spin and charge in La$_2$FeMnO$_6$ mixed perovskite

Nana Li[1, #], Fengren Fan[2, 3, #], Fei Sun[1], Yonggang Wang[1], Yongsheng Zhao[1], Fengliang Liu[1, 4], Qian Zhang[1], Daijo Ikuta[5], Yuming Xiao[5], Paul Chow[5], Steve M. Heald[6], Chengjun Sun[6], Dale Brewe[6], Aiguo Li[7], Xujie Lü[1], Ho-kwang Mao[1, 8], Daniel I. Khomskii[9], Hua Wu[2, 3, *], and Wenge Yang[1, *]

[1]Center for High Pressure Science and Technology Advanced Research (HPSTAR), Shanghai 201203, China

[2]Laboratory for Computational Physical Sciences (MOE), State Key Laboratory of Surface Physics, and Department of Physics, Fudan University, Shanghai 200433, China

[3]Collaborative Innovation Center of Advanced Microstructures, Nanjing University, Nanjing 210093, China

[4]The State Key Laboratory of Surface Physics, Department of Physics, and Laboratory of Advanced Materials, Fudan University, Shanghai 200433, China

[5]High Pressure Collaborative Access Team (HPCAT), X-Ray Science Division, Argonne National Laboratory, Argonne, Illinois 60439, United States

[6]X-ray Science Division, Advanced Photon Source, Argonne National Laboratory, Argonne, Illinois 60439, United States

[7]Shanghai Synchrotron Radiation Facilities, Shanghai Institute of Applied Physics, Chinese Academy of Sciences, Shanghai 201204, China

[8]Geophysical Laboratory, Carnegie Institution of Washington, Washington DC 200015, United States

[9]II Physikalisches Institut, Universität zu Köln, Zuelpicher Str. 77, Köln, 50937, Germany

[#] N.N.L. and F.R.F. contributed equally to this work.

*Correspondence and requests for materials should be addressed to yangwg@hpstar.ac.cn (W.Y.) and wuh@fudan.edu.cn (H.W.).

**Abstract:**

Spin crossover plays a central role in the structural instability, net magnetic moment modification, metallization, and even in superconductivity in corresponding materials. Most reports on the pressure-induced spin crossover with a large volume collapse so far focused on compounds with single transition metal. Here we report a comprehensive high-pressure investigation of a mixed Fe-Mn perovskite $La_2FeMnO_6$. Under pressure, the strong coupling between Fe and Mn leads to a combined valence/spin transition: $Fe^{3+}(S = 5/2) \rightarrow Fe^{2+}(S = 0)$ and $Mn^{3+}(S = 2) \rightarrow Mn^{4+}(S = 3/2)$, with an isostructural phase transition. The spin transitions of both Fe and Mn are offset by ~ 20 GPa of the onset pressure, and the lattice collapse occurs in between. Interestingly, $Fe^{3+}$ ion shows an abnormal behavior when it reaches a lower valence state ($Fe^{2+}$) accompanied by a + 0.5 eV energy shift in Fe $K$-absorption edge at 15 GPa. This process is associated with the charge-spin-orbital state transition from high spin $Fe^{3+}$ to low spin $Fe^{2+}$, caused by the significantly enhanced $t_{2g}$-$e_g$ crystal field splitting in the compressed lattice under high pressure. Density Functional Theory calculations confirm the energy preference of the high-pressure state with charge redistribution accompanied by spin state transition of Fe ions. Moreover, $La_2FeMnO_6$ maintains semiconductor behaviors even when the pressure reached 144.5 GPa as evidenced by the electrical transport measurements, despite the huge resistivity decreasing 7 orders of magnitude compared with that at ambient pressure. The investigation carried out here demonstrates high flexibility of double perovskites and their good potentials for optimizing the functionality of these materials.

# I. INTRODUCTION

After the first reports of the spin crossover (SCO) in transition metal ions, the interest and research in the SCO situation have shown rapid increase caused by its important physical significance and the interesting potential applications. [1-4] SCO materials play an important role in exploring new type spintronic devices which have been considered as a promising route to revolutionizing current logic and memory technologies [5-8]. For example, the memory effect of the SCO in [Fe(trz)$_3$](BF$_4$)$_2$ has been demonstrated [9, 10], showing their potential as switching elements in spintronic devices. SCO mainly occurs in the $3d$ magnetic systems ($3d^n$, $4 \leq n \leq 7$) where transitions occur between different spin states – high spin (HS) and low spin (LS), which leads to strong modification of the magnetic, electronic, optical and other properties of corresponding systems [9, 11-13]. In SCO materials, the interplay of several degrees of freedom including charge, spin, orbital, lattice, and a close proximity of different energy scales among them leads to their unique physical properties and can bring about intriguing functionalities, such as high-$T_c$ superconductivity, multiferroicity, etc [14-17].

As one of the most powerful methods to influence and study these phenomena, the pressure effect on SCO was proposed in the 1980s, and this method was used extensively since then [18-28]. For example, the pressure-induced spin-state transition in Fe$_x$O affects its compressibility [18], shear velocities [19], chemical stoichiometry [19], and electronic properties [20].

Among all SCO materials, members of the perovskite family ABO$_3$ (B = transition metal) attract most attentions because of their rich and versatile behavior, including excellent thermoelectric, photovoltaic, multiferroic properties [29-31], and consequently, leading to broad industry applications. Pressure is an effective tool of tuning crystal structure and electronic configuration of different materials. Especially on the transition metal (TM) site, pressure can significantly alter their behavior [32-34]. More interestingly, two TMs (B and B') can be introduced in perovskites, leading to the formation of double perovskites (dPv) A$_2$BB'O$_6$, with ordered or disordered B and B' sites. In such dPv we can have a rich selection on TMs with quite different electronic response to pressure, and one may then expect some novel properties. So far, no high pressure research has been reported on the coupling of two TM in such systems. Exploring the possible collaborative or competitive behavior of two TMs under pressure would provide a fundamental understanding of their structural and electronic properties, including possible phase transitions. Here, we study mixed (disordered) perovskite (mPv) La$_2$FeMnO$_6$, which at ambient pressure is ferrimagnetic [35], with Mn and Fe ions having quite different magnetic moments. Some high pressure works have been reported on the spin transition in single Fe or Mn perovskites. For instance, LaFeO$_3$ shows a Fe$^{3+}$ HS ($S$ = 5/2) to LS ($S$ = 1/2) transition accompanied by an antiferromagnetic to a non-magnetic transition in the broad pressure range of 30 to 50 GPa [36, 37]. In the layered perovskite CsMnF$_4$, the spin-crossover transition on Mn$^{3+}$ ($S$=2$\rightarrow$ $S$=1) takes place at 37 GPa with the suppression of the Jahn-Teller effect [38]. In the case of La$_2$FeMnO$_6$, both Fe and Mn could undergo pressure-induced spin-state transition which may also lead to structural instability. Particularly, upon compression the magnetic moment of transition-metal ions can be abruptly reduced. But besides that, the charge redistribution between Mn and Fe site could occur depending on the interplay of the lattice, spin, orbit, and valence.

Here, with a range of *in-situ* high-pressure techniques, including synchrotron X-ray diffraction (XRD), X-ray emission spectroscopy (XES), X-ray absorption near edge structure (XANES), and electrical transport (ET) measurements, we explore the correlation among crystal structure, electron spin, valence and electrical transport properties of $La_2FeMnO_6$. We demonstrate that in this system high pressure indeed induces both the valence change and the spin state transition occurring in a nontrivial way. The density functional theory calculations were also performed to further understand the mechanism of charge-spin-orbital state transitions and their interplay.

## II. DETAILS ON EXPERIMENTS AND CALCULATIONS
### A. Sample preparation

Sol-gel method was utilized to prepare $La_2FeMnO_6$. $La_2O_3$ (99.99%), MnO (99.99%), $Fe(NO_3)_3 \cdot 9H_2O$ (99.9%), and citric acid (AR) were used as raw materials. $La_2O_3$ and MnO were first dissolved in nitric acid with 1:1 ratio forming La and Mn nitrates and then diluted with distilled water. $Fe(NO_3)_3 \cdot 9H_2O$ was added to the distilled water and mixed with the La and Mn nitrates in a stoichiometric ratio of La:Fe:Mn = 2:1:1. Citric acid was then added as a fuel to the above solution to yield a citrate/nitrate ratio of 1.2. The mixed solution was continuously stirred with a magnetic agitator. The solution was further evaporated at 353 K until brown, sticky gel was formed. Subsequently, the gel was dried at 423 K. At last, the dried gel was calcined at 1273 K in air for 10 hours, followed intermediate grinding and pelletizations. X-ray diffraction measurements confirmed the final products as the pure *Pnma* structure.

### B. *In-situ* high pressure characterizations

Two sets of in-situ high-pressure XRD measurements were performed at the beamline 16-BM-D at the Advanced Photon Source (APS), Argonne National Laboratory (ANL), and beamline 15U at Shanghai Synchrotron Radiation Facility (SSRF). Symmetric diamond anvil cells (DAC) with anvil culet sizes of 300 $\mu$m and 100 $\mu$m, and rhenium gaskets were used. Neon was used as the pressure medium, and pressure was determined by the ruby luminescence method in the lower-pressure experiment [39], and Pt (111) *d*-spacing in the high pressure region [40]. Rietveld refinements on crystal structures at various pressures were performed using the General Structure Analysis System (GSAS) and graphical user interface EXPGUI package [41].

High pressure resistivity measurements were performed with a system consisting of a Keithley 6221 current source, a 2182A nanovoltmeter, and 7001 voltage/current switch system. Symmetric diamond anvil cells with 100-$\mu$m culet anvils were used; and a cubic boron nitride (*c*-BN) layer was inserted between metal gaskets and electrical leads. Four gold wires were arranged to contact the sample in the chamber for resistivity measurements as shown in Supplementary Material [42]. Pressure was calibrated by using the ruby luminescence method [40].

The high pressure XES measurements for Fe-$K_\beta$ and Mn-$K_\beta$ were conducted at 16-ID-D beamline at APS, ANL. To minimize the air scattering and absorption, helium pipes were placed at both incident and emission X-ray paths. Symmetric diamond anvil cells with 300-$\mu$m culet sized anvils were used with neon pressure medium. Beryllium gaskets were pre-compressed to 40-$\mu$m thick and drilled with a 150-$\mu$m-diameter hole as sample chambers. Pressure was calibrated by the ruby luminescence method [39].

The high pressure XANES of Fe $K$-edge was conducted at 20-BM-B beamline at APS, ANL. Two ionization chambers for $I_0$ (pre-sample intensity) and $I_1$ (post-sample intensity) and a focused X-ray beam were utilized for XANES measurements. A scan on the standard iron foil was performed for reference. A pair of 300-$\mu$m culet sized nano-diamond anvils was used in the DAC for the high pressure XANES. No pressure medium was used. Pre-compressed rhenium gasket, loading sample, and pressure calibration are the same as high pressure XRD experiments.

## C. DFT calculations

We chose a $\sqrt{2}\times\sqrt{2}\times2$ supercell with the Fe-Mn checkerboard arrangement for the DFT calculations. Another choice of a different structure turns out to give a same charge-spin-orbital transition under pressure, see **the DFT section in** Supplemental Material (Figure S1 [42]). All structures were relaxed by using the Vienna Ab-initio Simulation Package (VASP) [43] with Perdew-Burke-Ernzerhof (PBE) functional [44]. The energy cutoff was set as 400 eV. A $3\times3\times3$ Monkhorst-Pack k mesh was used. The electronic structures were calculated by using the full-potential augmented plane-wave plus local orbital method [45], in a local density approximation (LDA) [45]. The muffin-tin sphere for La, Fe, Mn, and O are 2.8, 2.0, 2.0, and 1.4 Bohr, respectively. The energy cutoff was set to be 14 Ryd. 250 k points were used for the energy integration over the whole Brillouin zone. Considering the electronic correlations, the Coulomb repulsion was included by LDA+U scheme [46] with the typical value of Hubbard U = 5.0 (4.0) eV and Hund exchange J = 1.0 (1.0) eV for Fe (Mn) $3d$ states. Our PBE+U calculations turn out to give almost the same results to the LDA+U ones, see the DFT section in Supplemental Material.

## III. RESULTS AND DISCUSSION
### A. Crystal structure evolution at high pressure

At ambient conditions, $La_2FeMnO_6$ crystallizes in an orthorhombic structure (space group *Pnma*, Z = 4) that has a three-dimensional network of corner-sharing $MnO_6$ and $FeO_6$ octahedrons with La atoms occupying the A-sites [47]. The Mn and Fe atoms are distributed randomly at the B sites. Synchrotron angle-dispersive XRD was utilized to study the structural evolution under high pressures. The analysis of one-dimensional profiles from two runs with different X-ray wavelengths is shown in Supplemental Material (Figure S2-Figure S6 [42]). Figure 1(a) displays the XRD patterns collected at selected pressures. All XRD data were analyzed with Rietveld refinement with GSAS software package [41]. Up to 87.8 GPa, the highest pressure studied in this work, the structure remained in its orthorhombic phase but had a noticeable volume collapse of 2.8% at pressure 28 to 45 GPa, which implies a first-order structural transition. The pressure dependence of the lattice parameters and the unit cell volume are displayed in Figure 1(b) and (c).

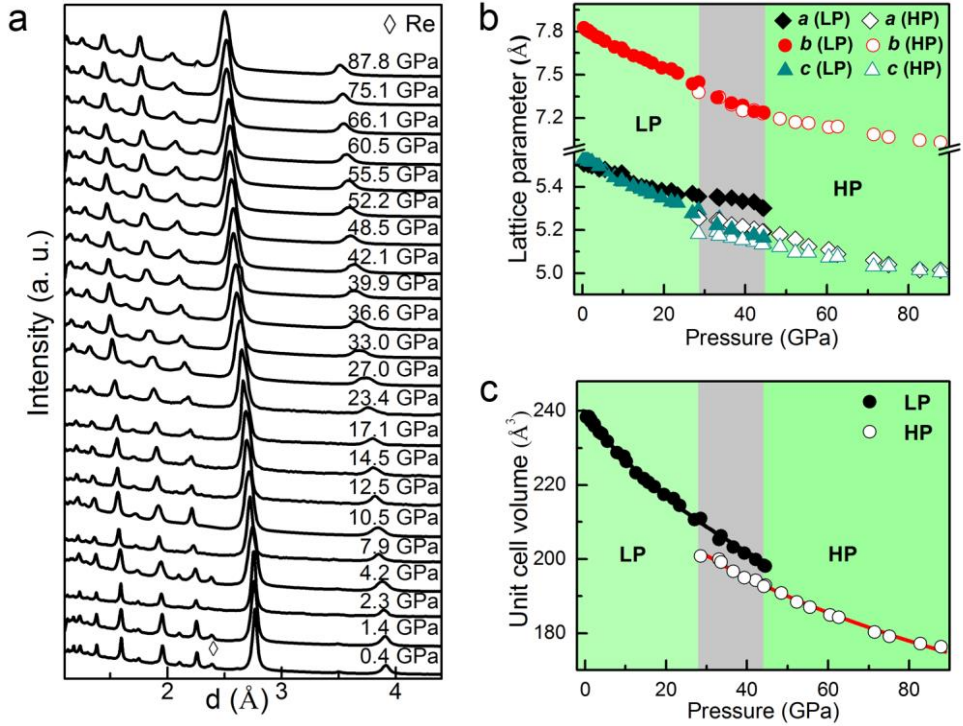

**FIG. 1.** Structural evolution of La$_2$FeMnO$_6$ under high pressure probed by synchrotron X-ray diffraction. (a) Selected angle dispersive XRD patterns of La$_2$FeMnO$_6$ as a function of pressure up to 87.8 GPa at room temperature. The corresponding changes in lattice parameters (b) and unit cell volume (c) in the entire pressure range. A volume discontinuity starting from 28 GPa can be seen clearly. The black and red solid lines in (c) refer to the fittings of the third-order Birch-Murnaghan equation of state in low pressure and high pressure phases respectively.

Two sets of P-V data were fitted separately in the pressure range below 28 GPa and above 45 GPa by using a third-order Birch-Murnaghan equation of state [48]. We obtained bulk modulus $B_0 = 164(3)$ GPa and its pressure derivative $B'_0 = 4.2(1)$ with $V_0 = 239.3(1)$ Å$^3$ for the low pressure (LP) phase; and $B_0 = 193(14)$ GPa and $B'_0 = 4.5(2)$ with $V_0 = 226.2(2)$ Å$^3$ for high pressure (HP) phase. There is a considerable increase in the bulk modulus along with a 2.8% volume collapse from LP to HP phase transition. Attempts to fit the HP patterns with a higher symmetry space group produced worse fitting results. We can certainly conclude that La$_2$FeMnO$_6$ undergoes a first-order isostructural phase transition in the 28-45 GPa range. This is quite a different compression behavior in comparison with LaFeO$_3$ and LaMnO$_3$ perovskites. LaFeO$_3$ transforms from orthorhombic to tetragonal structure around 28 to 50 GPa while LaMnO$_3$ stays in orthorhombic structure up to 40 GPa [49, 50]. The different compression behavior of La$_2$FeMnO$_6$ should be related to the distortion of (Fe/Mn)O$_6$ affected by the interaction between Mn and Fe atoms which will induce the competition between $FeO_6$ and $MnO_6$ under high pressure.

To gain insight into the distortion changes of (Fe/Mn)$O_6$, we further derived the (Fe/Mn)-O bond lengths and (Fe/Mn)-O-(Fe/Mn) bond angles from the Rietveld refinements, as shown in Figure 2(a) and Figure 2(b). Figure 2(c) displays the atomic arrangement of the unit cell with *Pnma* space group. In the range of 0 to 5 GPa, the two in-plane (Fe/Mn)-O2 bond lengths change in the opposite direction (one up and one down) while the out of plane (Fe/Mn)-O1 bond length increases gradually as pressure increases. This results in larger octahedron distortion. The increasing distortion is also reflected by the deviation away from 180 degrees of the (Fe/Mn)-O-(Fe/Mn) bond angles. From 5 GPa to 28 GPa, the (Fe/Mn)$O_6$ octahedron distortion is largely restored back to almost distortion free octahedron. Originating from the competition between $FeO_6$ and $MnO_6$, the average (Fe/Mn)$O_6$ octahedron shows two different distortion behaviors in different pressure regions. As the oxygen scattering power is much weaker than the rest metal elements (La, Fe and Mn), the uncertainty of oxygen positions from the Rietveld refinement is estimated and the error bars are added in the bonding length TM-O and angle TM-O-TM as shown in Figure 2(a) and Figure 2(b).

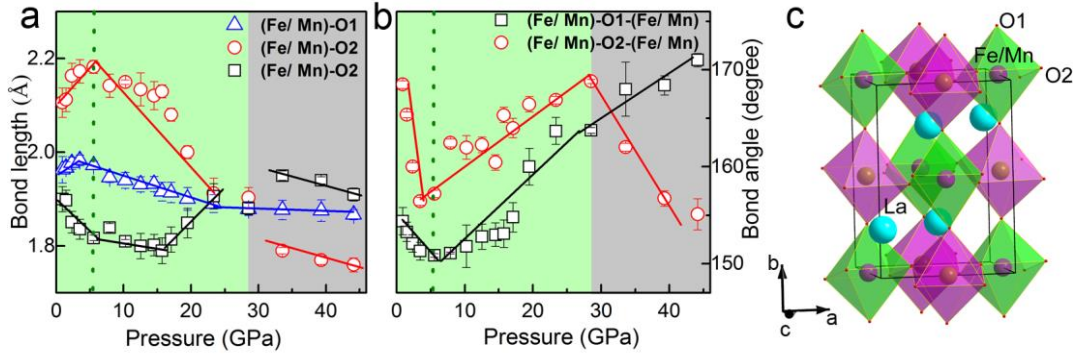

**FIG. 2.** The detailed atomic bond lengths and angles in $La_2FeMnO_6$ under high pressure. Bond length of (Fe/Mn)-O (a) and bond angle of (Fe/Mn)-O-(Fe/Mn) (b) as a function of pressure. (c) The crystal structure of the $La_2FeMnO_6$ unit cell with space group *Pnma*. In (a) and (b), the green shadows show the low pressure phase and the gray shadows show the mixed phase of high pressure phase and low pressure phase; the solid lines are used for the guide of the eye.

## B.  Spin state evolutions at high pressures

Pressure-induced increase in crystal field splitting can largely affect the spin configuration and thus minimize the total energy [51]. XES has been widely utilized to probe the spin state of transition metals. We conducted the XES on both Fe and Mn elements of $La_2FeMnO_6$ at various high pressures. The pressure dependent $K_{\beta1,3}$ and $K_{\beta'}$ emission spectra of Fe and Mn are shown in Figure 3(a) and (c) respectively. All spectra are normalized to the integrated area. The $K_\beta$ emission originates from the transition of the $1s$ core hole from a $3p$ level [52, 53]. Due to a net magnetic moment ($\mu$) effect on the $3d$ valence shell [54, 55], the $K_\beta$ emission spectrum is split into the main line $K_{\beta1,3}$ and a satellite line $K_{\beta'}$. The satellite intensity of $K_{\beta'}$ is proportional to the net spin of the $3d$ shell of the transition metal [56-58].

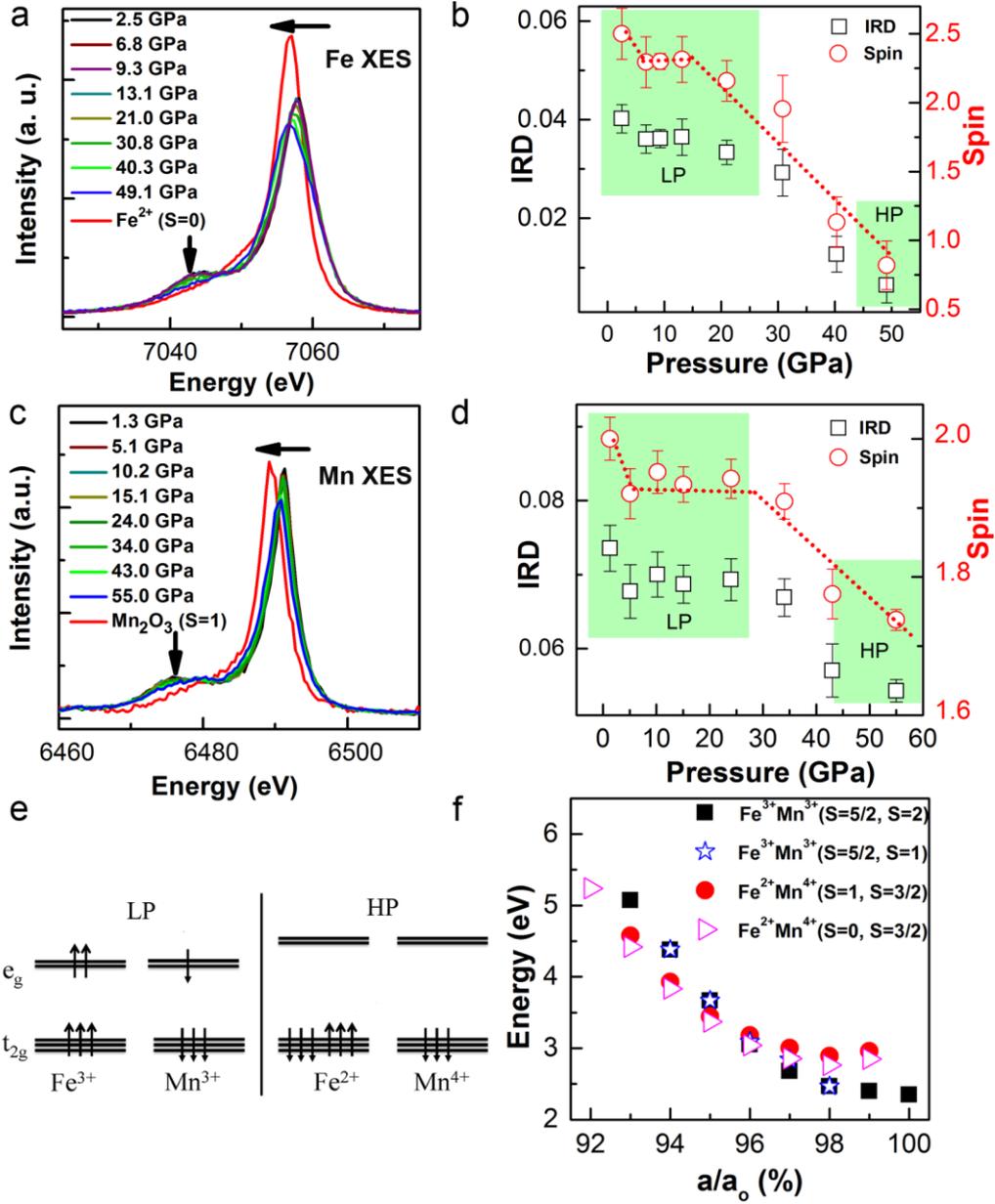

**FIG. 3.** Spin states of both Fe and Mn show a sluggish HS to LS transition under high pressure. (a) The XES spectrum of Fe. (b) The IRD and spin values of Fe under high pressure. (c) The XES spectrum of Mn. (d) The IRD and spin values of Mn under high pressure. (e) The $3d$ level diagrams of high spin $Fe^{3+}$ and $Mn^{3+}$ at LP and low spin $Fe^{2+}$ and $Mn^{4+}$ at HP. (f) Relative energy of the four stable states in a differently shrunk lattice in $La_2FeMnO_6$.

As shown in Figure 3(a) and 3(c), the $K_{\beta1,3}$ peak position is shifted to the lower energy, and the intensity of the $K_{\beta'}$ decreases for both Fe and Mn as pressure increased. We used the integrated relative difference (IRD) methods to process the XES data [59]. The details are described in Supplemental Material (Figure S7 [44]). The starting material has both $Fe^{3+}$ ($S = 5/2$) and $Mn^{3+}$ ($S = 2$) ions in their HS states. Then at various pressures, one can fit the IRD data of Fe and Mn by linearly interpolating the values for high and low spins from references [59]. Figure 3(b) and 3(d) show the IRD and spin values of Fe and Mn at different pressures. Both IRD and spin values have a sharp decrease around 6.8 GPa and again at 21.0 GPa for Fe. The same pattern was observed for Mn at 5.1 GPa and 34.0 GPa. Although these observations indicated a significant loss of magnetic moment in $La_2FeMnO_6$, neither Fe nor Mn transforms to the low spin state completely as the small shoulders from $K_{\beta'}$ peaks remained in both XES spectra to the highest pressure we reached. It is interesting to note that the spin values were stable in the range of 6.8-13.1 GPa for Fe and 5.1-24.0 GPa for Mn, where the crystal structure remained in the low pressure phase. Thus, there is a strong correlation between the crystal structure and spin states in $La_2FeMnO_6$. At ambient pressure, $Fe^{3+}$ in $La_2FeMnO_6$ has a $3d^5$ configuration. The five electrons occupy $d_{xy}$, $d_{xz}$, $d_{yz}$, $d_{x^2-y^2}$ and $d_{z^2}$ orbitals in a HS configuration based on Hund's rule. The net spin magnetic moment is mainly controlled by the competition between the crystal-field splitting $\Delta_{cf}$ (favorite for the LS state) and the intra-atomic Hund's exchange term $J$ (favorite for the HS state), but only $\Delta_{cf}$ is sensitive to external pressure. For $Fe^{3+}$ ($3d^5$) at HS, Hund energy gain is $-10J_H$ ($C_5^2 = 10$); for $Fe^{3+}$ at LS, Hund energy gain is $-4J_H$ (here $C_3^2 + C_2^2 = 4$) and crystal-field energy gain is $-2\Delta_{cf}$. Therefore, a rough estimation of the critical values of parameters for HS-LS transition is $\Delta_{cf} \approx 3J \sim 3$ eV which needs high pressure so that $\Delta_{cf}$ become larger. This explains the small decrease in the spin values of Fe and Mn when external pressure is below 6 GPa (Figure 3b and 3d). In the pressure range from 5 to 28 GPa, the distortion of the $(Fe/Mn)O_6$ octahedron decreases as shown in Figure 2, which inhibits the further splitting of crystal field. As a result, we observed the near constant spin values in the range of 6.8-13.1 GPa for Fe and 5.1-24.0 GPa for Mn respectively from XES.

The second spin transition for Fe starts between 13.1 and 21 GPa, which is much earlier than the structural phase transition (P~28 GPa), while the noticeable spin transition from Mn starts between 34 and 41 GPa, which is much later than the pressure of the structural transition. In general, Fe in $REFeO_3$ (RE = Rare earth) perovskite shows a sharp HS to LS transition accompanied with a structural phase transition with a sizeable volume collapse (3% for the $LaFeO_3$ case) above 30 GPa [37, 50]. The sluggish spin transition of Fe in $La_2FeMnO_6$ mixed perovskite at lower pressures may originate from the gradual valence/charge redistribution between Fe and Mn under pressure. It is known that $Fe^{2+}$ ions can be much easier transformed into a LS state than $Fe^{3+}$ [60]. As estimated above, the HS-LS transition for $Fe^{3+}$ occurs when crystal field splitting $\Delta_{cf} = 10$ Dq exceeds $3J_H$ [60]; whereas the same transition for $Fe^{2+}$ occurs at smaller crystal-field splitting $\Delta_{cf} = 2J_H$; i.e. at a lower pressure ($Fe^{2+}$ HS vs LS: $-10J_H$ vs $-6J_H-2\Delta_{cf}$). Such charge transfer, $Fe^{3+} + Mn^{3+} \rightarrow Fe^{2+}(LS) + Mn^{4+}$, may be induced by pressure because both LS $Fe^{2+}$ and $Mn^{4+}$ are smaller than the initial HS $Fe^{3+}$ and $Mn^{3+}$. In other words, increasing pressures can stabilize this new state. The $3d$ level diagrams of high spin $Fe^{3+}$ and $Mn^{3+}$ at LP and low spin $Fe^{2+}$ and $Mn^{4+}$ at HP are shown in Figure 3(e).

To verify the above-mentioned hypotheses, we carried out DFT calculations. We used the experimental ambient lattice constants and shrink the lattice from $a/a_0 = 1$ to 0.92 in steps of 0.01 to simulate the pressure effect, and we optimize the internal atomic positions. For efficient structural optimizations, we have used the VASP to do atomic relaxations on each crystal structure. Then we calculated their respective electronic structures to study the pressure effect and the possible valence-spin-orbital state transition using Wien2k. Table I listed the sum values of the ionic sizes in the different charge-spin states presumably presented in the pressure-induced transitions. In principle, the sum value should decrease as pressure increases in response to the shrinking lattice, and therefore this list would help us to recognize different charge-spin states.

**TABLE I. Ionic radii of Fe and Mn in different charge and spin states.**

| Fe | | Mn | | Radius sum (Å) |
|---|---|---|---|---|
| Charge-spin state | Ionic radii (Å) | Charge-spin state | Ionic radii (Å) | |
| +3 HS | 0.645 | +3 HS | 0.645 | 1.29 |
| +3 HS | 0.645 | +3 LS | 0.58 | 1.225 |
| +3 LS | 0.55 | +3 HS | 0.645 | 1.195 |
| **+3 LS** | **0.55** | **+3 LS** | **0.58** | **1.13** |
| +2 HS | 0.78 | +4 HS | 0.53 | 1.31 |
| **+2 LS** | **0.61** | **+4 HS** | **0.53** | **1.14** |

Through density functional theory calculations, we have found that out of the above listed six states, four have stable solutions, as seen in Figure 3(f). It is clear that $Fe^{3+}$(HS)-$Mn^{3+}$(HS) is the most stable state at ambient pressure ($a/a_0 = 1$) while the $Fe^{2+}$(LS)-$Mn^{4+}$(HS) becomes most robust, the lowest energy state at high pressure ($a/a_0 = 0.92$). It is interesting that the $Fe^{2+}$(HS)-$Mn^{4+}$(HS) state, which has the largest sum value of the ionic sizes comparing to the $Fe^{3+}$(HS)-$Mn^{3+}$(HS) state, is unstable at ambient pressure in our calculations and converges exactly to the $Fe^{3+}$(HS)-$Mn^{3+}$(HS) ground state. On the other hand, the $Fe^{3+}$(LS)-$Mn^{3+}$(LS) state, which has the smallest sum value of the ionic size, is unstable at the high pressure and converges to the $Fe^{2+}$(LS)-$Mn^{4+}$ state. Considering a most plausible charge fluctuation and the induced intermediate excited $Fe^{2+}$/$Mn^{4+}$ state, a Fe-O-Mn superexchange will yield an antiferromagnetic (AF) coupling between HS $Fe^{3+}$ and HS $Mn^{3+}$. In addition to a large Hund exchange splitting of the HS $Fe^{3+}$ ($t_{2g}^3 e_g^2$, $S = 5/2$) and the Jahn-Teller crystal field splitting of HS $Mn^{3+}$ ($t_{2g}^3 e_g^1$, $S = 2$), electron correlations will determine the ferrimagnetic behavior at the ambient pressure (AP) for $La_2FeMnO_6$. In contrast, $La_2FeMnO_6$ transforms into the $Fe^{2+}$ (LS)-$Mn^{4+}$ state under high pressure (HP). Then the LS $Fe^{2+}$ has a closed $t_{2g}^6$ shell and is nonmagnetic ($S = 0$), while $Mn^{4+}$ has a closed $t_{2g}^3$ subshell ($S = 3/2$). This partially explains why the $Fe^{2+}$(LS) and $Mn^{4+}$(LS) state are more stable than the $Fe^{3+}$ (LS)-$Mn^{3+}$ (LS) state that has a common open $t_{2g}$ shell although both states have very similar small ionic sizes. The closed-shell LS $Fe^{2+}$/$Mn^{4+}$ state well matches the compact structure of $La_2FeMnO_6$ at high pressure. Because LS $Fe^{2+}$ ($S = 0$) is nonmagnetic and the magnetic $Mn^{4+}$ ions ($S = 3/2$) are diluted, HP $La_2FeMnO_6$ could be weakly antiferromagnetic or even paramagnetic.

The induced HS to LS transition is a gradual process for Fe atoms starting at a low pressure and extending over a broad pressure range. Due to the random occupation of Fe and Mn ions at B site, this valence and spin transition only happens in a suitable local environment. Only when the average spin of Fe reaches a critical value at 28 GPa, it triggers total lattice instability and induces a sluggish phase transition. However, at the onset pressure of phase transition, the $Mn^{3+}$ ions remain in a HS state as their crystal-field did not yet increase sufficiently. The opposite trends of (Fe/Mn)-O-(Fe/Mn) bonding angles and the separation of (Fe/Mn)-O bonding lengths beyond 28 GPa illustrated in Figure 2(a) and Figure 2(b), indicates that the further compression intensifies the distortion of $(Fe/Mn)O_6$ octahedron. When the pressure reaches to the range of 34-40 GPa, which is at least 6 GPa higher than the on-set structural transition pressure of 28 GPa, we start to observe a spin state change on Mn. In this process, the Mn undergoes a valence change from $Mn^{3+}$ to $Mn^{4+}$, which corresponds to the spin state transition from $S = 2$ ($t_{2g}^3 e_g^1$) to $S = 3/2$ ($t_{2g}^3 e_g^0$). This matches the trend obtained from the IRD analysis. In addition to the IRD analysis, the XES measurement results suggested that the Fe spin is below 1.0 while Mn spin is near 1.7 at the highest pressure (P~50 GPa). With further increasing pressure till lattice shrinkage of about $a/a_0$ ~ 0.92 (P~80 GPa), as follows from the DFT calculations shown in Figure 3(f), one can expect that the final spin state will reach $S = 0$ for Fe and $S = 3/2$ for Mn.

## C. The valence of Fe at high pressures

XANES is a sensitive tool to probe the valence state of elements. Compared to the Fe $K$-edge XANES profile at ambient pressure, we noticed that the main absorption edge shifted to higher energy side by 0.5 eV at 15 GPa (Figure 4a). In theory, this chemical shift indicates that $Fe^{3+}$ ions try to reach a higher valence state (e.g, $Fe^{4+}$). However, this contradicts our conclusion drawn from the first principles calculations, where $Fe^{3+}$ tends to obtain one electron from $Mn^{3+}$ thus having $Fe^{2+}$ and $Mn^{4+}$. To clarify this inconsistency, we studied the charge-spin-orbital state transition of the Fe ions under pressure in detail by applying DFT method. As shown in Figure 4(b), the Fe $3d$ and $4p$ density of states (DOS) have a relative energy shift between the ambient pressure and high pressure phases. Our calculations indicate that the HP LS $Fe^{2+}$ has a smaller energy separation between the $1s$ core level and the Fermi level (chemical potential) than that of AP HS $Fe^{3+}$ by 3.1 eV, which is in line with the above common knowledge about the chemical shift. Therefore, when we plot the $3d$ and $4p$ DOS of the AP HS $Fe^{3+}$ with the Fermi level set at zero energy (Figure 4b top), we need to shift downwards the Fermi level of the HP phase and the corresponding DOS curves by 3.1 eV (Figure 4b bottom). Given that Fe $4p$ state is much more delocalized and has only a tiny DOS intensity, we need to trace it via the localized $3d$ state due to the $3d$-$4p$ hybridization. In the AP phase, the Fermi level lies in a tiny gap between the up and down spin $3d$ channels of the HS $Fe^{3+}$. However, in the HP phase, the Fermi level sits at the top of the valence band with a large energy gap between the occupied $t_{2g}^6$ of the LS $Fe^{2+}$ and the unoccupied $e_g^0$. This is the result of the large $t_{2g}$-$e_g$ crystal field splitting and electron correlations in the compressed HP lattice. Thus, the bottom of the conduction band in the HP phase is higher than that of the AP phase by 0.75 eV. This well explains the observed upward shift of 0.5 eV in Fe $K$-edge XANES under pressure. Based on the aforementioned analyses, we can conclude that the unusual upward shift of the Fe $K$-edge XANES is the result of the charge-spin-orbital state transition from the HS $Fe^{3+}$ to LS $Fe^{2+}$ under pressure, thus offering a counter chemical intuition but a correct physical picture.

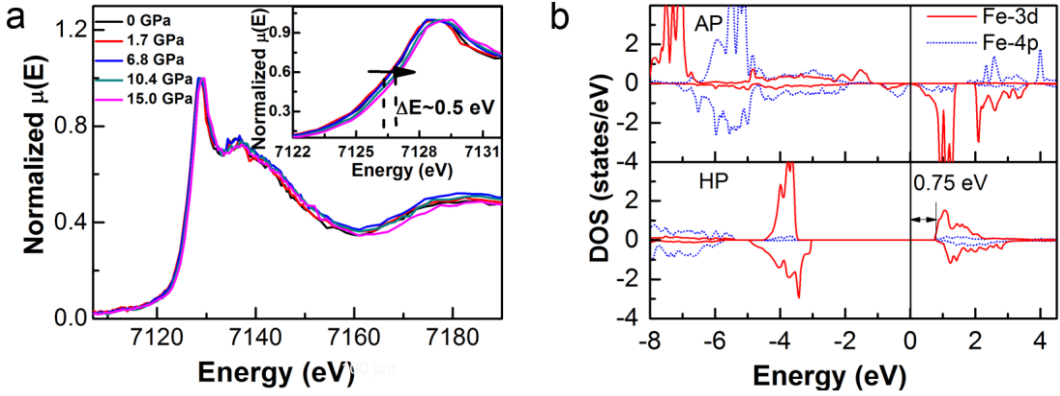

**FIG. 4.** The valence state of Fe in $La_2FeMnO_6$ under high pressure. (a) Normalized XANES spectra of Fe $K$-edge at different pressures. The inset shows the zoomed-in feature of the main absorption peaks. (b) DOS of Fe $3d$ and $4p$ orbitals at ambient pressure (AP) and high pressure (HP). The $4p$ DOS is magnified by 50 times for clarity.

### D. Electronic transport properties at high pressure

As pressure induces significant changes on the lattice, can induce the charge transfer between the two TM elements, and changes their orbital/spin configurations, one may expect a large change in the electronic properties [6, 16, 61-64]. To test the changes of the electronic properties of $La_2FeMnO_6$, we conducted the electrical resistivity measurements on $La_2FeMnO_6$ at room temperature and at low temperature (in the range from liquid nitrogen to room temperature) with pressure up to 144.5 GPa. The results are displayed in Figure 5(a) and 5(b). Overall, the resistivity decreases monotonically with pressure. Although the resistivity dropped by 7 orders of magnitude from ambient pressure to 144.5 GPa, its semiconductor behavior persists at the highest pressure as shown in Figure 5(b) and Figure S8 [42]. Interestingly, the decrease of resistivity accelerated between 20 and 50 GPa as shown in Figure 5(a). The sharp drop of resistivity indicated a possible electronic transition. Moreover, XRD detected that the decrease of resistivity is associated with the onsite isostructural transition starting from 20 GPa which affects the bandwidth of $e_g$ orbitals. In earlier research, the top of the valence band was shown to be dominated by the Mn $3d$ $e_g$ state [35]. In our case the $e_g$ orbital broadening decreases the band gap of $La_2FeMnO_6$, but not enough to close the gap to make the system metallic up to 144.5 GPa.

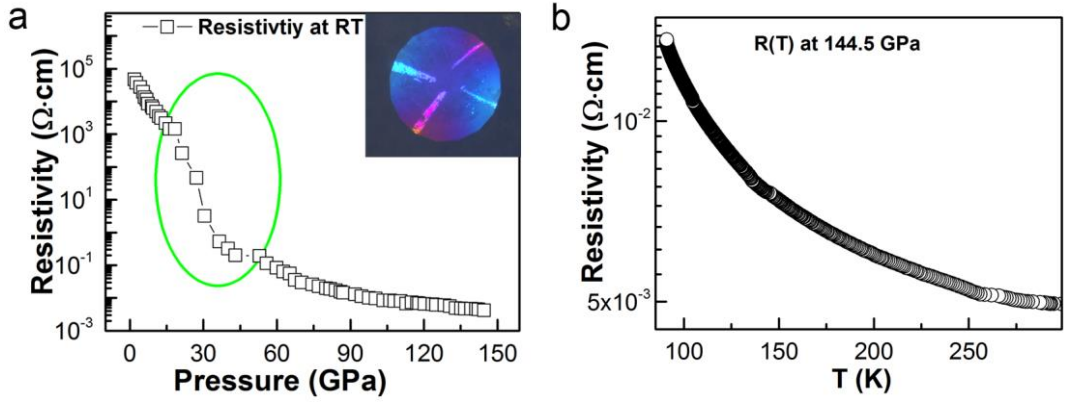

**FIG. 5.** The electric transport property of $La_2FeMnO_6$ under high pressure. (a) The electrical transport as a function of pressure. The inset in (a) is the diagram of paving electrodes for the electrical transport measurement. (b) The resistivity as a function of temperature at 144.5 GPa. Despite 7 orders of magnitude decrease in resistivity from ambient pressure to 144.5 GPa, the $La_2FeMnO_6$ remains a semiconductor. There is a sharp drop of resistivity between 20 and 50 GPa hatched by a green ellipse in (a).

## IV. CONCLUSIONS

In conclusion, by using different pressure response of two transition metals at B site of a double perovskite, we have successfully manipulated the structural and electronic properties of the mixed perovskite $La_2FeMnO_6$ under high pressure. Although the crystal structure of $La_2FeMnO_6$ remained the same from the ambient phase up to 87.8 GPa, the valence states of Fe and Mn ions have changed as the electrons are transferred from $Mn^{3+}$ to $Fe^{3+}$, accompanied by their spin and orbital reconfigurations to minimize the total energy of the system. Due to the random occupation of Fe and Mn in $La_2FeMnO_6$, both the structural and spin transition are rather broad, extending over 20 GPa range. The interplay of the lattice, spin and orbital degrees of freedom leads to both the charge redistribution between Fe and Mn ions, and to the spin-state transition of Fe ions. Despite the fact that the resistivity decreases by 7 orders of magnitude as compared with that at ambient pressure, $La_2FeMnO_6$ maintained its semiconductor behaviors even when the pressure reached 144.5 GPa. The current study demonstrates a great versatility of double perovskites and their great potential to unravel interesting interplay of different degrees of freedom in strongly correlated compounds. We can achieve in them some tailoring of intriguing properties by carefully selecting the element species at the B site and their ratio. Our study could provide a useful guidance for designing novel spintronic materials with desired properties.

## ACKNOWLEDGMENTS

This work was financially supported by the National Nature Science Foundation of China (Grant No. 51527801, U1530402, 11474059, 11674064), and by the National Key Research Program of China (Grant No. 2016YFA0300700). The work of D.Kh. was funded by the Deutsche Forschungsgemeinschaft, project number 277146847 - CRC 1238. HPCAT operations are supported by DOE-NNSA under Award No.

DE-NA0001974 and DOE-BES under Award No. DE-FG02-99ER45775, with partial instrumentation funding by NSF. The gas loading was performed at GeoSoilEnviroCARS, APS, ANL, supported by EAR-1128799 and DE-FG02-94ER14466. APS is supported by DOE-BES, under Contract No. DE-AC02-06CH11357.